\documentclass{article}
\usepackage{amsmath,graphicx,url}

\title{Acoustic event detection for multiple overlapping similar sources}

\author{Dan Stowell and David Clayton \\
Queen Mary University of London\\
     London, UK \\
     dan.stowell@qmul.ac.uk}

\begin{document}

\maketitle

\begin{abstract}
Many current paradigms for acoustic event detection (AED) are not adapted to the organic variability of natural sounds,
and/or they assume a limit on the number of simultaneous sources:
often only one source, or one source of each type, may be active.
These aspects are highly undesirable for applications such as bird population monitoring.
We introduce a simple method modelling the onsets, durations and offsets of acoustic events
to avoid intrinsic limits on polyphony or on inter-event temporal patterns.
We evaluate the method in a case study with over 3000 zebra finch calls.
In comparison against a HMM-based method we find it more accurate at recovering acoustic events,
and more robust for estimating calling rates.
\end{abstract}


\section{Introduction}
\label{sec:intro}

Acoustic event detection (AED) is useful for various purposes, such as security monitoring, wildlife monitoring, and music transcription \cite{Giannoulis:2013b,Stowell:2015,Stiefelhagen:2007,Marques:2012}.
Many approaches to AED assume that the acoustic scene is \textit{monophonic}---having no simultaneous or overlapping events---which is unrealistic but useful for some applications.
Approaches which allow for polyphonic scenes are more flexible,
but often assume that each stream is different in kind, allowing one monophonic stream for each class of event considered \cite{Diment:2013,Ewert:2015}. They may also assume a fixed number of simultaneous streams, for example when source separation is applied as a first step and then each separated channel is treated as a monophonic scene \cite{Heittola:2011}.

In this paper we explore approaches for AED in cases with an unknown number of similar sources. As an example, consider a sound recording in which a flock of birds can be heard calling, all of the same species.
This is representative of practical scenarios in which AED might assist
ecologists or conservation organisations wishing to estimate the total number of \textit{individuals} detected in a recording, or alternatively the total number of \textit{calls} in a recording. Both such ``point counts'' are used (in most cases manually detected) for monitoring trends in breeding populations \cite{Marques:2012}.

Note the specific information need: for overall event counts,
which is distinct from the ``transcription'' task in which the objective is to recover a list of the true events.
An exact transcription would itself give us an exact value for the event count;
but an imperfect transcription may be an inefficient or biased route to count estimation.
In the abstract, event counting tasks are regression problems, and may not require an estimated event transcript at any point.
Also important is that we wish to avoid placing limits or inappropriate biases on the number of events that can be simultaneously active.
Methods that assume monophonic sequences per event class are likely to be inappropriate, and other methods may bias estimation due to assumptions implicit in their models.
With this in mind, we briefly consider some event detection paradigms in previous work.


To decompose an audio scene,
one strand of research uses non-negative matrix factorisation (NMF),
in particular convolutive NMF which allows events to have spectro-temporal structure
\cite{OGrady:2006}.
However, these models are inflexible about the temporal evolution within an event,
depending on good matching of spectro-temporal templates.
This is particularly problematic for sounds with much inherent within-class variability such as animal calls.

Hidden Markov models (HMMs) have been used in various systems for acoustic event detection (e.g.\ \cite{Benetos:2012,Diment:2013}).
These can allow for variation in the temporal evolution of events.
However a typical HMM corresponds to a monophonic model of events;
extensions such as the factorial HMM extend this to a specific fixed number of parallel sources,
and thus retain strong limits on the level of polyphony.
In one example of polyphonic adapatation of HMM tracking, \cite{Diment:2013}
train and apply a standard HMM for event detection, where in their case each state corresponds to a class of event.
To achieve polyphonic detection they perform multiple Viterbi decoding passes:
after each Viterbi pass, the states used are taken out of consideration for the subsequent passes.
In this way a transcription is obtained which allows multiple event classes to occur in parallel.
However it does not allow multiple simultaneous events of the same class,
and retains the fixed limit on polyphony.

In Section \ref{sec:hmm} we will describe an alternative way to adapt a HMM to multiple detection scenario.
However, that is primarily as a point of comparison against the main model we wish to explore here,
which uses an \textit{onset-duration-offset} model of acoustic events to allow for unbounded polyphony.
We describe this method in the next section.
Then we will describe our alternative HMM method, before evaluating both methods using a dataset of bird calls.

\section{Onset-duration-offset model}
\label{sec:odo}

\begin{figure}[t]
  \centering
  \centerline{\includegraphics[width=0.8\columnwidth]{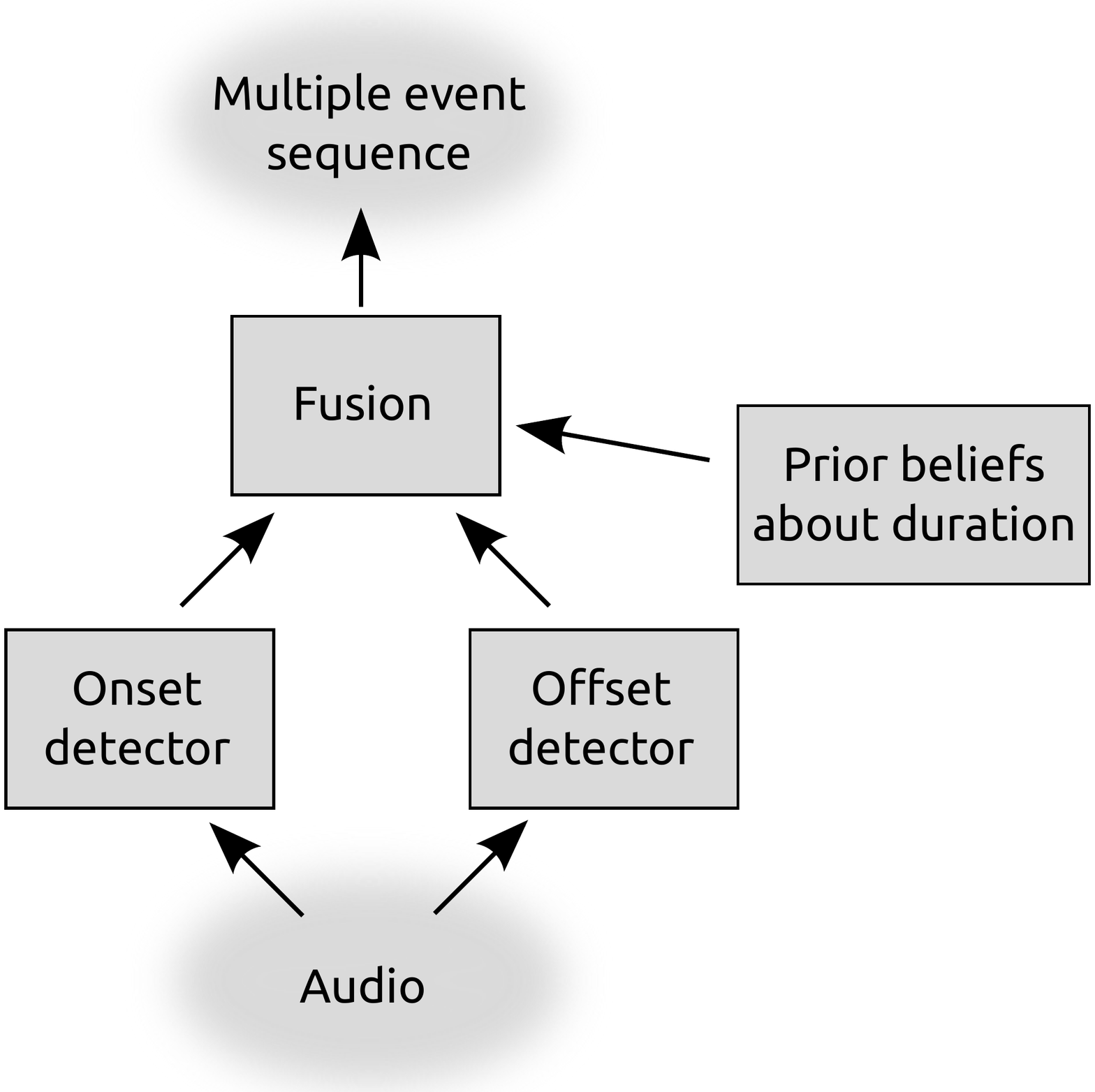}}
  \caption{Schematic diagram of onset-duration-offset event detection model.}
  \label{fig:odoscheme}
\end{figure}

Physiological studies indicate that biological auditory processing involves early-stage ``edge detectors'' having separate auditory detection units for onsets and for offsets,
both in humans \cite{Chait:2008} 
and in songbirds \cite{Woolley:2009}.  
The information from these detectors is then combined in later processing for cognition of ``auditory objects'' or events.
Although there is no requirement for our computational systems to mimic the organisation found in nature,
this suggests that a processing strategy starting with onset and offset detectors and combining their outputs may be fruitful.
We can combine onsets and offsets with other information to yield posterior beliefs about the events observed (Figure \ref{fig:odoscheme}).
If these components are based on the characteristics of individual events, and not the temporal relationships between events,
we should be able to design a system that imposes few constraints or biases on the observable event patterns.

The scheme just presented assumes that the onset and offset characteristics are the reliable, relatively invariant characteristics of the events of interest.
However, it makes no strong assumptions on event durations, nor even the signal content in the middle of the event, thus allowing for organic variability.
It also makes no strong assumptions on the temporal occurrence patterns, and in particular the level of polyphony is unbounded:
at any particular time, if the system overall has detected $k$ more onsets than offsets, then the current number of parallel active events is $k$.

Our approach is probabilistic: we will use onset/offset detectors that yield detection probabilities at each time point,
and our prior beliefs about event duration will be expressed as a distribution over durations.
To combine these probabilities together, we characterise acoustic events in a two-dimensional space indexed by \textit{onset time} $t$ and \textit{duration} $\tau$.
We characterise the conditional probability of an event at some point in that space as
\begin{equation*}
p_\text{evt}(t, \tau | \mathbf{y}) \propto p_\text{on}(t | \mathbf{y}) \, p_\text{off}(t + \tau | \mathbf{y}) \, p_\text{dur}(\tau)
\end{equation*}
where $\mathbf{y}$ is the observation (the audio signal).
The conditional probabilities $p_\text{on}$ and $p_\text{off}$ come from the detectors, and $p_\text{dur}$ is our duration prior.
When dealing with discretised time (as we do here), this Bernoulli model imposes a mild constraint: no two events can have exactly the same onset time and duration.
This constraint is very mild, since events can co-occur in our scheme as long as they have slight mutual differences in onset time \textit{or} duration.
(We will impose a slightly stronger constraint to recover an event transcript, described shortly.)

\begin{figure}[t]
  \centering
  \centerline{\includegraphics[width=0.8\columnwidth,clip,trim=0mm 40mm 0mm 5mm]{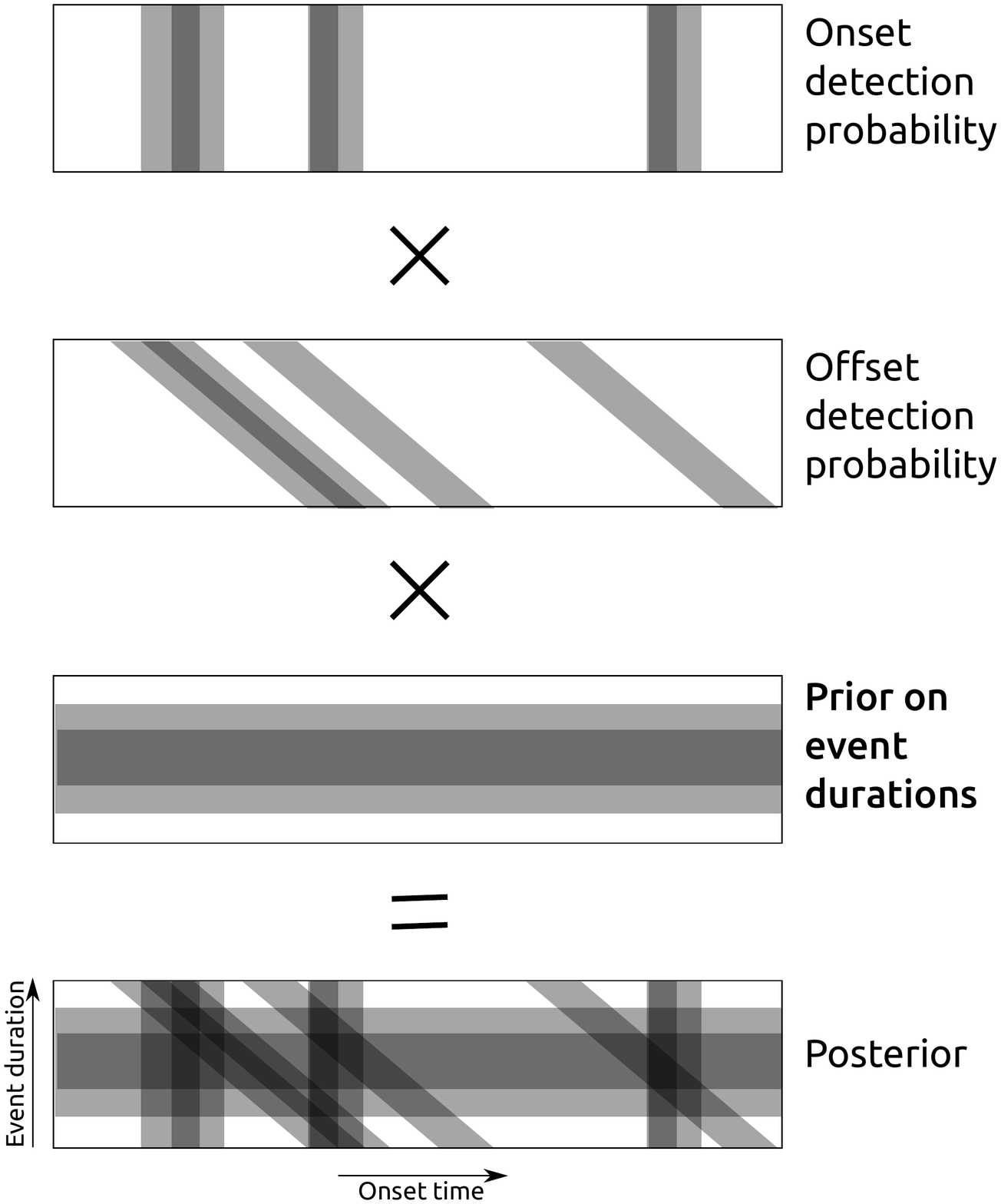}}
  \caption{Schematic diagram of how probabilities are combined in the onset-duration-offset event detection model. Each source of probabilistic information is a marginal with respect to a different direction in the [onset $\times$ duration] space.}
  \label{fig:onsetdurposterior}
\end{figure}

Each of our probabilistic sources of information (onsets, offsets, durations) gives us information that acts as a one-dimensional marginal, when considered in our two-dimensional [onset $\times$ duration] space (Figure \ref{fig:onsetdurposterior}).
Note particularly that offset detection probabilities are translated into [onset $\times$ duration] space with an off-axis influence, since offset time is equivalent to onset time plus duration.
In this space, we assume that the three types of probability information are conditionally independent and multiply them together to produce a posterior ``intensity'' over possible events
(Figure \ref{fig:onsetdurposterior}).
This could be thresholded to give a polyphonic event sequence,
or marginalised to give posterior beliefs about onsets and offsets.
Such posterior beliefs will be related to the raw onset/offset detections but refined using the other information sources.
Note that the posterior is not a probability distribution:
it does not sum up to one over our 2D space.
It represents a set of binomial probabilities; the sum over the 2D posterior gives the expected number of events.

In the implementation we use here, for onset and offset detectors we use random forest regression (cf.\ \cite{Phan:2014}).
Our detectors take spectrogram patches as input, the spectrograms having been treated with background noise reduction by median-thresholding, and then first differencing in time.
The trained random forest outputs detection probabilities for each patch.
Since the regression makes an independence assumption for adjacent (overlapping) patches,
the outputs are liable to be correlated in time;
to reduce this effect, after training the random forest we then train an ordinary least squares regression from a sliding window of 11 outputs from the detector onto the ground truth, to recover ``sharper'' detections.
To implement our prior on event durations, we will train a Gaussian mixture model (GMM) on the durations observed in training data.

We note some resemblances between our approach and that of \cite{Phan:2014}.
Those authors also use random forest regression as a recognition component that contributes towards an eventual event segmentation.
However, their method is fundamentally different in that its elements for recognition are not the onsets/offsets,
but the frames ``within'' an event, which have been augmented with pointers to their associated onset/offset.
For this and other reasons their approach is limited to monophonic event detection.

In the following we will refer to our onset-duration-offset model as \textit{ODO} for short.

\subsection{Recovering an event transcript vs.\ event counts}

To recover a definite set of events, our 2D posterior can be thresholded using a threshold determined during the training phase.
In practice, however, we observe that this tends to yield a large number of duplicated events, since any particular ground-truth event of duration $\tau$ will often be detected with a relatively strong probability for duration $\tau+1$, $\tau-1$, etc., each of which is a separate position in our 2D posterior.
This effect can be reduced by imposing further assumptions: perhaps about the maximum polyphony, or the temporal pattern.
In the present work we wish to avoid imposing assumptions that might strongly bias event counts and the like.
We choose to impose an assumption which is partly implicit in the detectors themselves:
the assumption that only one onset, and one offset, may happen in any time frame.
This assumption may bias detection in very dense audio recordings,
but for many densities encountered in practice this assumption holds almost always.
Hence to recover an event transcript, we keep only the events whose posterior probability is stronger than all other events with matching onset time or offset time.
From these events we then select a threshold for discarding low-probability events.

Taking Figure \ref{fig:onsetdurposterior} as an example, in the posterior we see that the density due to the second detected onset overlaps in 2D space with two possible offsets.
In this case we would keep no more than one of these events, namely that which yields the strongest probability.


A straightforward way to count events is to count the number of items included in an event transcript.
However, as just described, producing an event transcript requires making ``hard'' thresholding decisions, discarding some information from the posterior.
We can avoid the transcription step and simply use the sum of the posterior, over the time range of interest, as the expected event count in that time range.
We will use this in our evaluation.

\section{Hidden Markov model for event count}
\label{sec:hmm}

As an alternative to the method presented above, we also describe a hidden Markov model (HMM) method for event detection.
As described in Section \ref{sec:intro}, HMM approaches to event detection impose limits on the possible polyphony,
and also may bias the durations and timing patterns of detected events.
With those caveats acknowledged, we wish to use the HMM as a point of comparison since it is in widespread use.
So in order to detect events of a single type, but with potential polyphony,
we apply a HMM but where the hidden states are not simply `on' and `off',
but the count of currently-active events, i.e. the count of events that have started but not yet ended.
The state space is thus $\{0, 1, 2 ... K\}$ where $K$ is the maximum number of simultaneous events observed in the training data.

For modelling the observations, we train a separate Gaussian mixture model (GMM) with ten components for each cardinality.
Again we use spectral patches to train this model, but without differencing them in time since in this case we aim to model states rather than transitions.

To recover an event transcription from this HMM, we perform Viterbi decoding.
From the decoded sequence of cardinalities, we deduce onset and offset times,
and we associate onsets and offsets with each other in order of occurrence.
This transcript is then also used for event counting.

\subsection{Combining the two models}
\label{sec:combined}

The ODO and HMM models we have described offer two very different approaches to event detection.
We note that it is possible to combine the two, as follows.
We can expand the HMM state space to include not only the current event cardinality,
but also two binary indicators of whether the current frame includes an onset and/or an offset.
This expands the set of HMM states by a factor of four.
Not all state transitions are possible:
e.g.\ a change in cardinality from 3 to 4 can only occur when the onset state is 1.
We do not impose such limits manually but allow the system to learn them from the transitions seen in training data.

Even with this expanded state space, HMM-based models inherit the limitations already mentioned:
cardinalities higher than seen in the training data will not be correctly detected,
and the HMM may bias timing patterns.
However in the following evaluation we will compare the empirical characteristics of the methods we have described.

\section{Evaluation}
\label{sec:eval}


We recorded a set of four female zebra finches (\textit{Taeniopygia guttata}) in an indoor aviary.
The birds exchanged contact calls at a rate of approx 40 calls per minute.
Birds were recorded for extended periods, and their calls were transcribed as event sequences.
Transcription was performed separately by two human annotators,
whose annotations were combined automatically,
and any discrepancies resolved by the first author.

This recording setup was designed for use in various studies;
in the present paper we use it as a case study for event detection.
We took two 30-minute recording sessions, recorded with the same birds but on separate days,
and used these two sessions for two-fold crossvalidation.
The 30-minute sessions contained 1663 and 1770 annotated calls in total.
Here we use single-channel omni mic recordings of the sessions.

The true polyphony in the original recordings ranged from zero to four.
In order to investigate heavier densities, as well as to investigate the effect of density mismatches between training and test data,
for each 30-minute session we also created an artificial 10-minute mixture with the three 10-minute segments superimposed.
All experiments thus used the same set of calls, but in some cases the training or test data was ``folded'' down to a denser 10-minute recording by superimposition.

As in other event detection evaluations \cite{Giannoulis:2013b,Stowell:2015},
for evaluating event transcription we use the F-measure metric
and we consider an event to be correctly recovered if the onset matches within a fixed tolerance
and the duration matches within $50\%$ of the true duration.
The typical event duration in this data was approx 100 ms, so we chose $\pm 25$ ms as our tolerance.

Separately, for evaluating event counts we divide the data into ten-second windows and measure the RMS error between the true and estimated number of events for each window.
Note that both systems do exhibit some miscalibration, in that their estimated counts even on the training data could exhibit a multiplicative deviation from the truth.
For the ODO system we believe this is largely due to the independence assumption already mentioned in the underlying detectors, and thus more sophisticated edge detectors might remedy this.
To account for this most basic aspect of miscalibration, during training of all systems we used the training data to choose a multiplicative calibration factor to apply to all event counts.
Calibration did not make use of test data.
RMS error statistics are reported from the calibrated outputs.


We tested the following five configurations of event detector:

\begin{itemize}
	\item
	The full ODO system of Section \ref{sec:odo}.
	\item
	The ODO system with a flat duration prior rather than the GMM. This constrained events to lie within a reasonable duration but did not learn a distribution on the duration, and allowed us to probe the extent of the benefit provided by the GMM duration prior.
	\item
	The HMM system of Section \ref{sec:hmm}.
	\item
	The combined ODO+HMM system of Section \ref{sec:combined}.
	\item
	The raw output from the onset-detector component. This can be evaluated for event counts only, not transcription, but indicates the extent to which the detector component contributes to ODO performance.
\end{itemize}

Specific implementation details were as follows.
Audio was recorded at 96 kHz, and spectrograms calculated with frame length 2048, 50\% overlap and Hann windowing.
Spectral information outside the range 0.5--20 kHz was discarded.
Noise reduction was implemented using spectral median-subtraction, with the median calculated in a sliding ten-second window for each spectral band.
Spectral patches for the onset/offset detectors were taken from the time-differenced spectrogram, of size five frames before and five frames after the frame under consideration.
The detectors were implemented using random forest regression from the sklearn module, with 20 trees.
Spectral patches for the HMM-GMM modelling were taken from the non-time-differenced spectrogram, of size five frames after the frame under consideration.

\begin{figure}[t]
  \centering
  \centerline{\includegraphics[width=0.99\columnwidth,clip,trim=0mm 0mm 0mm 0mm]{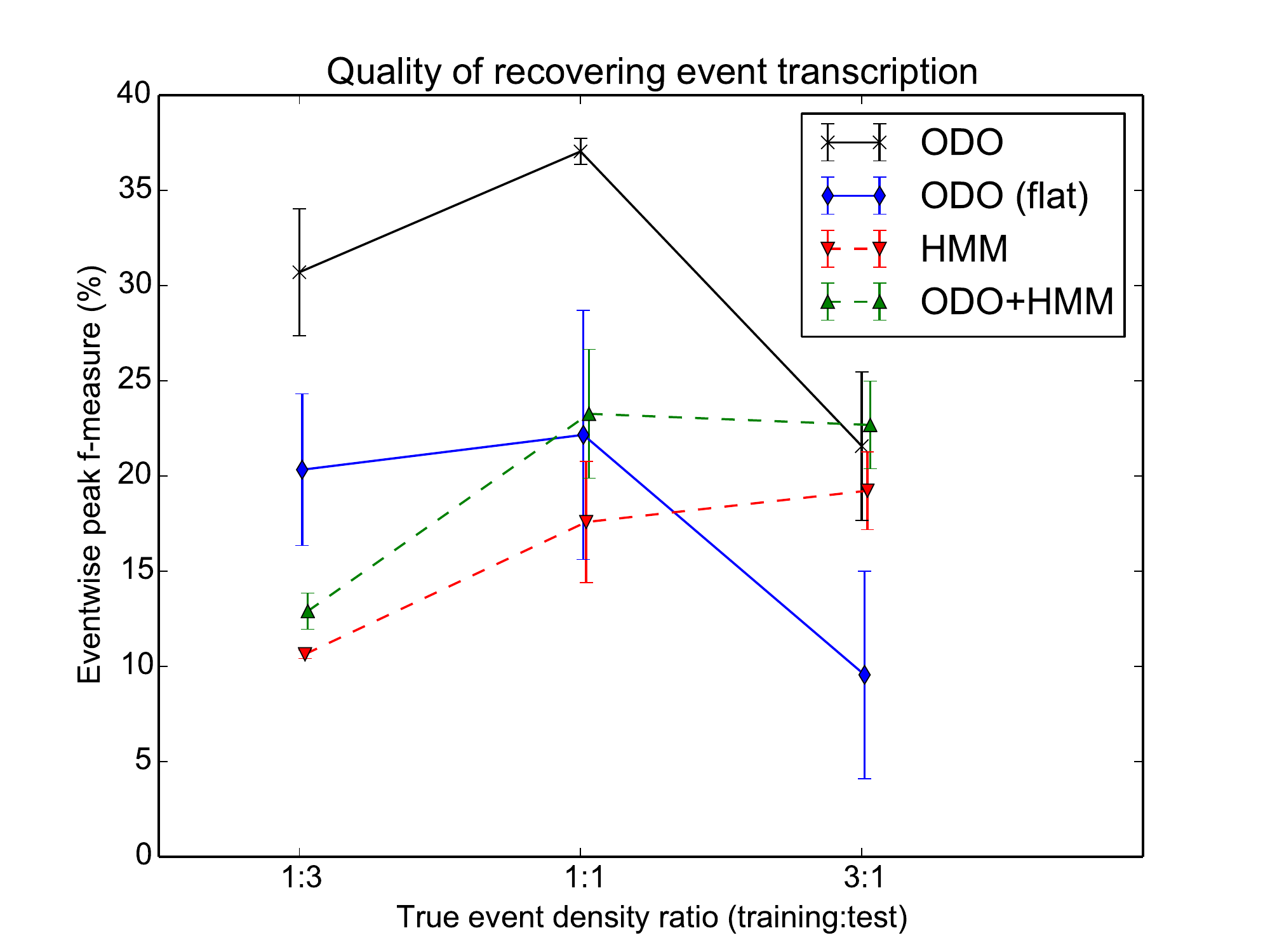}}
  \caption{Event transcription results (F-measure), averaged over the two crossvalidation folds. Error bars cover the range of results obtained within individual folds.}
  \label{fig:results_eventf}
\end{figure}

Results for event transcription (Figure \ref{fig:results_eventf}) show a number of tendencies.
Firstly the full ODO model consistently outperforms the ODO model with flat duration prior,
indicating that the learned prior on event durations adds useful information.
Secondly, the HMM system generally performs much worse than the ODO system.
This poor performance might be attributed to various points of difference between the two systems,
and so it is interesting to observe that the combined ODO+HMM system improves on HMM but does not approach ODO's strong performance,
despite making use of ODO's onset/offset output as part of its input observations.

The experiments with mismatched density in training and test give a general performance degradation for all systems,
indicating that there is still some way to go to perform detection robust to very wide variation in event density.
The HMM-based systems perform relatively well in the experiment with increased training density---an exception to the general pattern.
Conversely, the poor performance of the HMM-based systems in the experiment with increased testing density
matches expectations since the training did not encompass all the event cardinalities found in the testing data.

In best conditions, our ODO system achieved an F-measure of around 37\%.
Although quite distant from the ideal of 100\%, it is on a similar scale as the results reported for state-of-the-art methods for related tasks \cite{Giannoulis:2013b,Stowell:2015}.

\begin{figure}[t]
  \centering
  \centerline{\includegraphics[width=0.99\columnwidth,clip,trim=0mm 0mm 0mm 0mm]{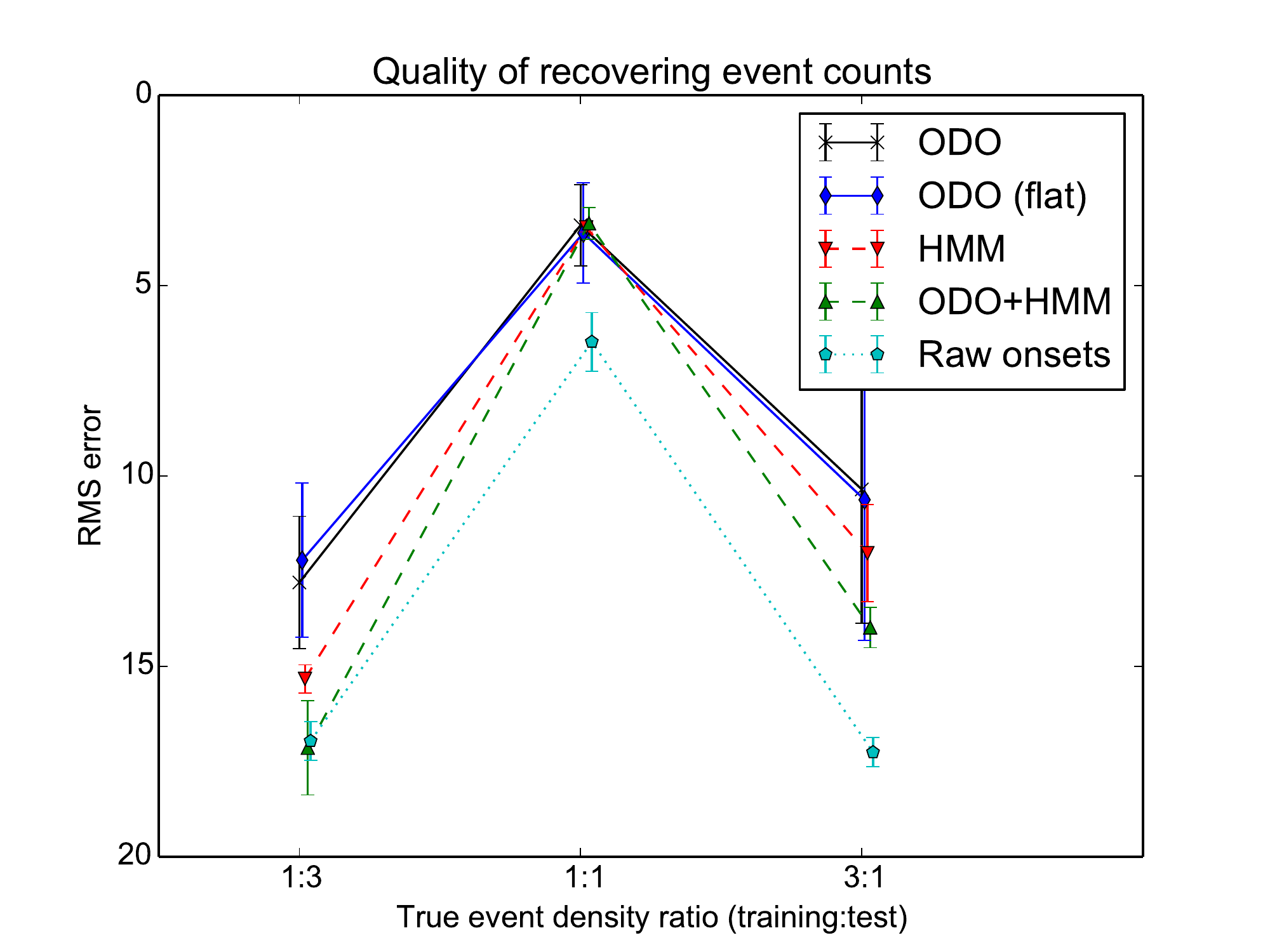}}
  \caption{Count estimation results (RMS error of counts in ten-second time windows). The y-axis is inverted so that upwards means better performance, to match Figure \ref{fig:results_eventf}.}
  \label{fig:results_countrms}
\end{figure}

Results for event counting (Figure \ref{fig:results_countrms}) show a slightly different picture.
Again, the mismatch in training and testing conditions has a general negative impact on performance.
In the main experiment, most of the systems perform at very similar quality levels.
This is except for the raw onset detector output, included for comparison, which performs notably worse than the full systems
---illustrating that the ODO method performs much better than its underlying detector.
However, although the HMM and ODO+HMM systems achieve similar performance as ODO in the main experiment,
this is not the case in the experiments with mismatched training and test densities, for which their performance degrades further.

Taken together, these evaluations indicate that the ODO method is more accurate and more robust than the HMM method
for detecting or counting events in polyphonic bird recordings such as those we have studied.
The method can be used for any data with events well-characterised by `landmarks' such as onsets and offsets, including animal and human sounds.
However we note that all the systems evaluated here showed quite some decay in performance when evaluated with mismatched event densities.
Improving these polyphonic detection paradigms to be robust to these wide ranges of event density remains as future work.
Good detector components must be a key to strong performance: for the present work we used simple detection methods using spectral patches as data;
improvements such as feature learning \cite{Stowell:2014b} could improve detection performance.
It also remains to evaluate systems on a wider range of event-annotated audio recordings.

\section{Acknowledgment}
\label{sec:ack}

Our thanks to Maeve McMahon for help with bird management,
and to Robert Jack and Alex Wilson for data annotation.

\clearpage
\bibliographystyle{IEEEtran}
\bibliography{../../refs}

\begin{thebibliography}{10}
\providecommand{\url}[1]{#1}
\csname url@samestyle\endcsname
\providecommand{\newblock}{\relax}
\providecommand{\bibinfo}[2]{#2}
\providecommand{\BIBentrySTDinterwordspacing}{\spaceskip=0pt\relax}
\providecommand{\BIBentryALTinterwordstretchfactor}{4}
\providecommand{\BIBentryALTinterwordspacing}{\spaceskip=\fontdimen2\font plus
\BIBentryALTinterwordstretchfactor\fontdimen3\font minus
  \fontdimen4\font\relax}
\providecommand{\BIBforeignlanguage}[2]{{%
\expandafter\ifx\csname l@#1\endcsname\relax
\typeout{** WARNING: IEEEtran.bst: No hyphenation pattern has been}%
\typeout{** loaded for the language `#1'. Using the pattern for}%
\typeout{** the default language instead.}%
\else
\language=\csname l@#1\endcsname
\fi
#2}}
\providecommand{\BIBdecl}{\relax}
\BIBdecl

\bibitem{Giannoulis:2013b}
D.~Giannoulis, E.~Benetos, D.~Stowell, M.~Rossignol, M.~Lagrange, and M.~D.
  Plumbley, ``Detection and classification of acoustic scenes and events: an
  {IEEE AASP} challenge,'' in \emph{Proceedings of the Workshop on Applications
  of Signal Processing to Audio and Acoustics (WASPAA)}, 2013.

\bibitem{Stowell:2015}
D.~Stowell, D.~Giannoulis, E.~Benetos, M.~Lagrange, and M.~D. Plumbley,
  ``Detection and classification of acoustic scenes and events,'' \emph{{IEEE}
  Transactions on Multimedia}, 2015.

\bibitem{Stiefelhagen:2007}
R.~Stiefelhagen, K.~Bernardin, R.~Bowers, J.~Garofolo, D.~Mostefa, and
  P.~Soundararajan, ``The {CLEAR} 2006 evaluation,'' \emph{Multimodal
  Technologies for Perception of Humans}, pp. 1--44, 2007.

\bibitem{Marques:2012}
T.~A. Marques, L.~Thomas, S.~W. Martin, D.~K. Mellinger, J.~A. Ward, D.~J.
  Moretti, D.~Harris, and P.~L. Tyack, ``Estimating animal population density
  using passive acoustics,'' \emph{Biological Reviews}, 2012.

\bibitem{Diment:2013}
A.~Diment, T.~Heittola, and T.~Virtanen, ``Sound event detection for office
  live and office synthetic {AASP} challenge,'' in \emph{IEEE AASP Challenge on
  Detection and Classification of Acoustic Scenes and Events (WASPAA 2013
  special session)}, 2013,
  http://c4dm.eecs.qmul.ac.uk/sceneseventschallenge/abstracts/OL/DHV.pdf.

\bibitem{Ewert:2015}
S.~Ewert, M.~D. Plumbley, and M.~Sandler, ``A dynamic programming variant of
  non-negative matrix deconvolution for the transcription of struck string
  instruments,'' in \emph{Proc ICASSP 2015}, 2015.

\bibitem{Heittola:2011}
\BIBentryALTinterwordspacing
T.~Heittola, A.~Mesaros, T.~Virtanen, and A.~Eronen, ``Sound event detection in
  multisource environments using source separation,'' in \emph{Workshop on
  Machine Listening in Multisource Environments (CHiME 2011)}, 2011, pp.
  36--40. [Online]. Available:
  \url{http://spandh.dcs.shef.ac.uk/projects/chime/workshop/papers/pS32_heittola.pdf}
\BIBentrySTDinterwordspacing

\bibitem{OGrady:2006}
P.~D. O'Grady and B.~A. Pearlmutter, ``Convolutive non-negative matrix
  factorisation with a sparseness constraint,'' in \emph{Machine Learning for
  Signal Processing, 2006. Proceedings of the 2006 16th IEEE Signal Processing
  Society Workshop on}.\hskip 1em plus 0.5em minus 0.4em\relax IEEE, 2006, pp.
  427--432.

\bibitem{Benetos:2012}
E.~Benetos, M.~Lagrange, and S.~Dixon, ``Characterisation of acoustic scenes
  using a temporally-constrained shift-invariant model,'' in \emph{Conference
  on Digital Audio Effects Conference (DAFx-12)}, vol.~17, 2012, p.~21.

\bibitem{Chait:2008}
M.~Chait, D.~Poeppel, and J.~Z. Simon, ``Auditory temporal edge detection in
  human auditory cortex,'' \emph{Brain research}, vol. 1213, pp. 78--90, 2008.

\bibitem{Woolley:2009}
S.~M. Woolley, P.~R. Gill, T.~Fremouw, and F.~E. Theunissen, ``Functional
  groups in the avian auditory system,'' \emph{The Journal of Neuroscience},
  vol.~29, no.~9, pp. 2780--2793, 2009.

\bibitem{Phan:2014}
H.~Phan, M.~Maasz, R.~Mazur, and A.~Mertins, ``Random regression forests for
  acoustic event detection and classification,'' \emph{IEEE/ACM Transactions on
  Audio, Speech, and Language Processing}, vol.~23, no.~1, pp. 20--31, 2014.

\bibitem{Stowell:2014b}
D.~Stowell and M.~D. Plumbley, ``Automatic large-scale classification of bird
  sounds is strongly improved by unsupervised feature learning,'' \emph{PeerJ},
  vol.~2, p. e488, 2014.

\end{thebibliography}

\end{document}